\documentclass[epj]{webofc}
\usepackage[utf8]{inputenc}
\usepackage[varg]{txfonts}   % Web of Conferences font
\usepackage{booktabs}
\usepackage{xcolor}
\definecolor{darkred}{rgb}{0.4,0.0,0.0}
\definecolor{darkgreen}{rgb}{0.0,0.4,0.0}
\definecolor{darkblue}{rgb}{0.0,0.0,0.4}
\usepackage[bookmarks,linktocpage,colorlinks,
    linkcolor = darkred,
    urlcolor  = darkblue,
    citecolor = darkgreen]{hyperref}
\usepackage{subfigure}
\usepackage[overlay,absolute]{textpos}
\wocname{EPJ Web of Conferences}
\woctitle{Lattice2017}
\usepackage{dsfont}
\newcommand{\id}{\mathds{1}}
\newcommand{\be}{\begin{equation}}
\newcommand{\ee}{\end{equation}}
\newcommand{\bea}{\begin{eqnarray}}
\newcommand{\eea}{\end{eqnarray}}

\newcommand{\nn}{\nonumber}
\newcommand{\Tr}{\text{Tr}}

\newcommand{\ii}{\text{i}\,}

\def\slash#1{#1\!\!\!/\,}
\def\Dslash{\slash D}

\begin{document}
%%%%%%%%%%%%%%%%%%%%%%%%%%%%%%%%%%%%%%%%%%%%%%%%%%%%%%%%%%%%%%%%%%%%%%%%%%%%%
%
\begin{flushright}
\footnotesize{
CERN-TH-2017-210 \\
CP$^3$-Origins-2017-44 DNRF90
}
\end{flushright}
\vspace*{-0.7cm}
\selectlanguage{english}
%----------------------------------------------------------------------------
\title{%
Alternatives to the stochastic ``noise vector'' approach
}
%----------------------------------------------------------------------------
\author{%
\firstname{Philippe} \lastname{de Forcrand}\inst{1,2}\fnsep\thanks{\email{forcrand@phys.ethz.ch}} \and
\firstname{Benjamin} \lastname{J\"ager}\inst{1,3}
% etc.
}
%----------------------------------------------------------------------------
\institute{%
Institute for Theoretical Physics, ETH, CH-8093 Z\"urich, Switzerland
\and
CERN, Theory Division, CH-1211 Geneva, Switzerland
\and
CP3-Origins \& Danish Institute for Advanced Study, Department of
Mathematics and Computer Science, University of Southern Denmark, 5230 Odense M, Denmark
}
%----------------------------------------------------------------------------
\abstract{%
Several important observables, like the quark condensate and the Taylor coefficients 
of the expansion of the QCD pressure with respect to the chemical potential, are based on the trace of 
the inverse Dirac operator and of its powers. Such traces are traditionally estimated 
with "noise vectors" sandwiching the operator. We explore alternative approaches based on polynomial 
approximations of the inverse Dirac operator.
}

%----------------------------------------------------------------------------
\maketitle
%----------------------------------------------------------------------------
\section{Introduction}\label{intro}

Measuring observables on an ensemble of gauge field configurations
can require computing resources similar to those required to generate
the configurations. However, comparatively little effort has been devoted
to optimizing the computing strategy for such measurements.
An important category of fermionic observables can be expressed
as the trace of some function of the Dirac operator, or some
other related operator. For example, the quark condensate
$\bar\psi \psi$ is obtained from the trace of the inverse of
the Dirac operator $\Dslash$. Generically, the trace for a given gauge field configuration is 
estimated by using ``noise vectors'' $\eta_k, k=1,..,n$, via

\be
 {\rm Tr} A  \approx \frac{1}{n} \sum_{k=1}^n
\eta_k^\dagger A \eta_k
\label{eq_trace}
\ee

\noindent
with $\langle \eta_k^{i\dagger} \eta_k^j \rangle = \delta^{ij}$.
Here, $i$ and $j$ label components of a noise vector $\eta$, and $\langle .. \rangle$
means  averaging over noise vectors (for a given gauge field).
Clearly, there are two sources of noise here: 
the ``{\em stochastic noise}'' coming from the limited number $n$ of noise vectors
used on a given gauge field configuration; 
and the ``{\em gauge noise}'' coming from the fluctuations caused by the variation of
the gauge field from configuration to configuration. 
The balance between these two types of noise depends on the observable,
and on the parameters of the Monte Carlo simulation: in particular, fluctuations coming from
the gauge field will severely increase in the vicinity of a phase transition.

Here, we consider two situations of relevance at finite temperature: \\
$(i)$ the measurement of the chiral condensate and its first 4 moments near the
finite-temperature deconfinement transition. The moments are necessary
to construct the Binder cumulant (or kurtosis), whose value at the phase transition is known by
universality; \\
$(ii)$ the measurement of the Taylor coefficients of the expansion of the
pressure in $\mu/T$. \\ 
In the latter situation, it is not unusual to take 1000 noise vectors or more per
gauge configuration. Can one be more efficient?

%----------------------------------------------------------------------------
\section{Related recent progress}\label{sec-1}

Applied mathematicians have proposed a novel approach to determine 
the spectrum of a Hermitian matrix
$A$~\cite{di2016efficient}.
Instead of obtaining the eigenvalues directly by a Krylov-based method,
the idea is to determine the ``eigenvalue count'', namely the {\em number}
$\mu(a,b)$ of eigenvalues in some interval $[a,b]$. Upon varying $a$ and $b$,
the positions of the jumps in the eigenvalue count give the desired eigenvalues.
Thus, one needs a sliding bandpass filter $f(x,a,b)$ which is $= 1$ if $x \in [a,b]$,
$0$ otherwise.
In practice: \\
 \hspace*{0.5cm} - The function $f$ is approximated by Chebyshev polynomials and applied to the
matrix $A$. Then, 
\hspace*{0.75cm} the ``eigenvalue count'' $\mu(a,b)$ is $\Tr~ f(A,a,b)$. \\
\hspace*{0.5cm} - The trace is approximated by using a small number of noise vectors following
eq.(\ref{eq_trace}). \\
One great advantage is that, by varying the coefficients of the polynomial,
$a$ and $b$ can be varied at will without any additional matrix-vector multiplication,
 and thereby the {\em complete} spectrum of $A$
can be determined at low cost.

Two lattice groups have tested this approach so far: \\
$\bullet$ In \cite{Fodor:2016hke}, a degree-8000 Chebyshev approximation was constructed and
applied to $\Dslash^\dagger \Dslash$, using 20 noise vectors for the trace.
This required $80000$ Dirac-matrix-vector multiplications. These are
very moderate resources, considering that the lattice size was $64^3\times 96$,
and that the whole spectrum could be extracted. After averaging over
10 gauge configurations, a precise determination of $\langle \bar\psi \psi \rangle$
could be obtained. \\
$\bullet$ In \cite{Cossu:2016eqs}, an overlap Dirac operator was used, on the same lattice size,
with a Chebyshev approximation of the same degree 8000, and 1 noise vector
per configuration only. The gauge ensemble consisted of 50 configurations.
Again, a precise determination of $\langle \bar\psi \psi \rangle$ could be
obtained.

These promising first results were obtained at zero temperature. We are interested
in applying a similar approach at $T \sim T_c$, the deconfinement temperature,
where the ``gauge noise'' is expected to be much greater. And we want,
in addition to $\bar\psi \psi = \sum_i \lambda_i^{-1}$, higher negative moments
of the eigenvalues, i.e. $\sum_i \lambda_i^{-m}, m=1,2,3,4,..$.

%----------------------------------------------------------------------------

\begin{figure}
\hspace*{-0.2cm}
\includegraphics[width=0.50\linewidth]{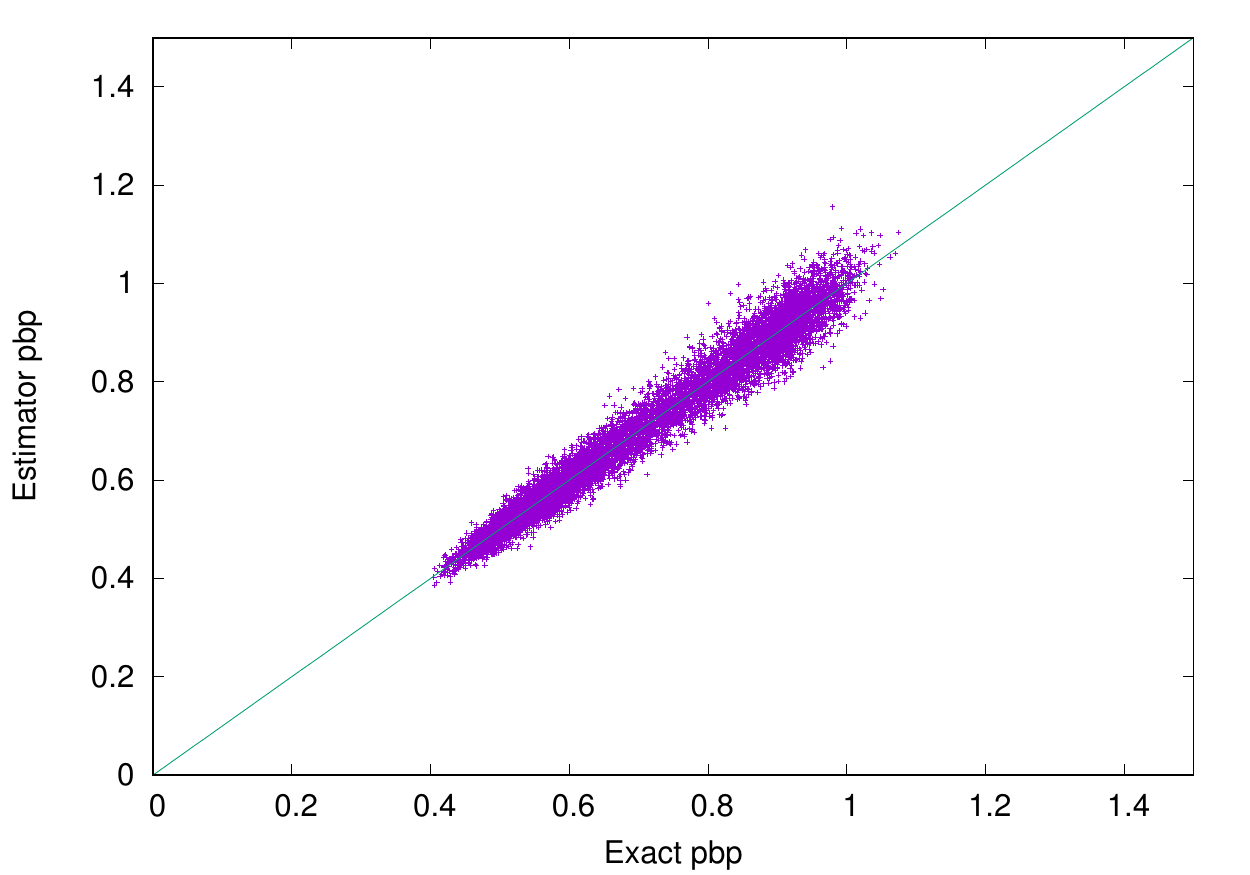}
\includegraphics[width=0.50\linewidth]{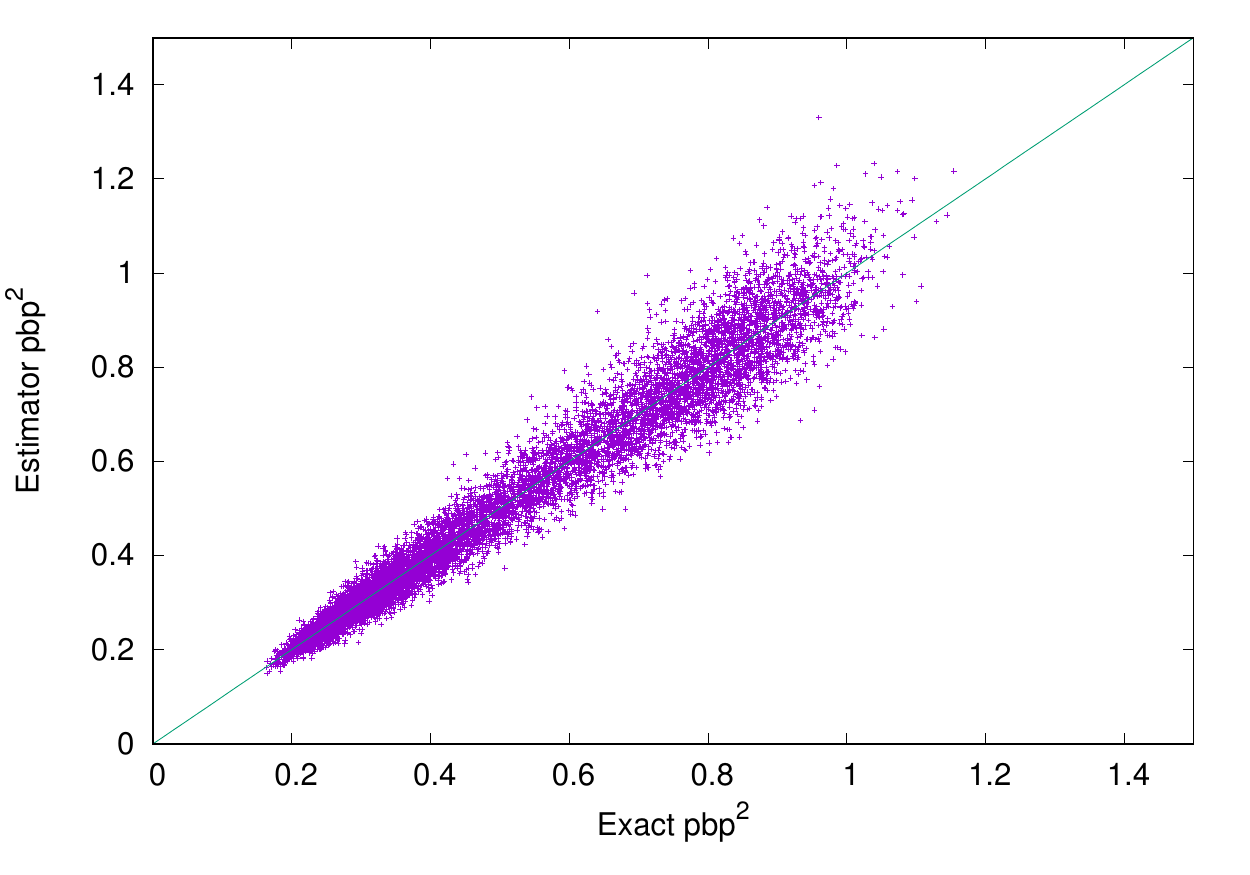} \\
\hspace*{-0.2cm}
\includegraphics[width=0.50\linewidth]{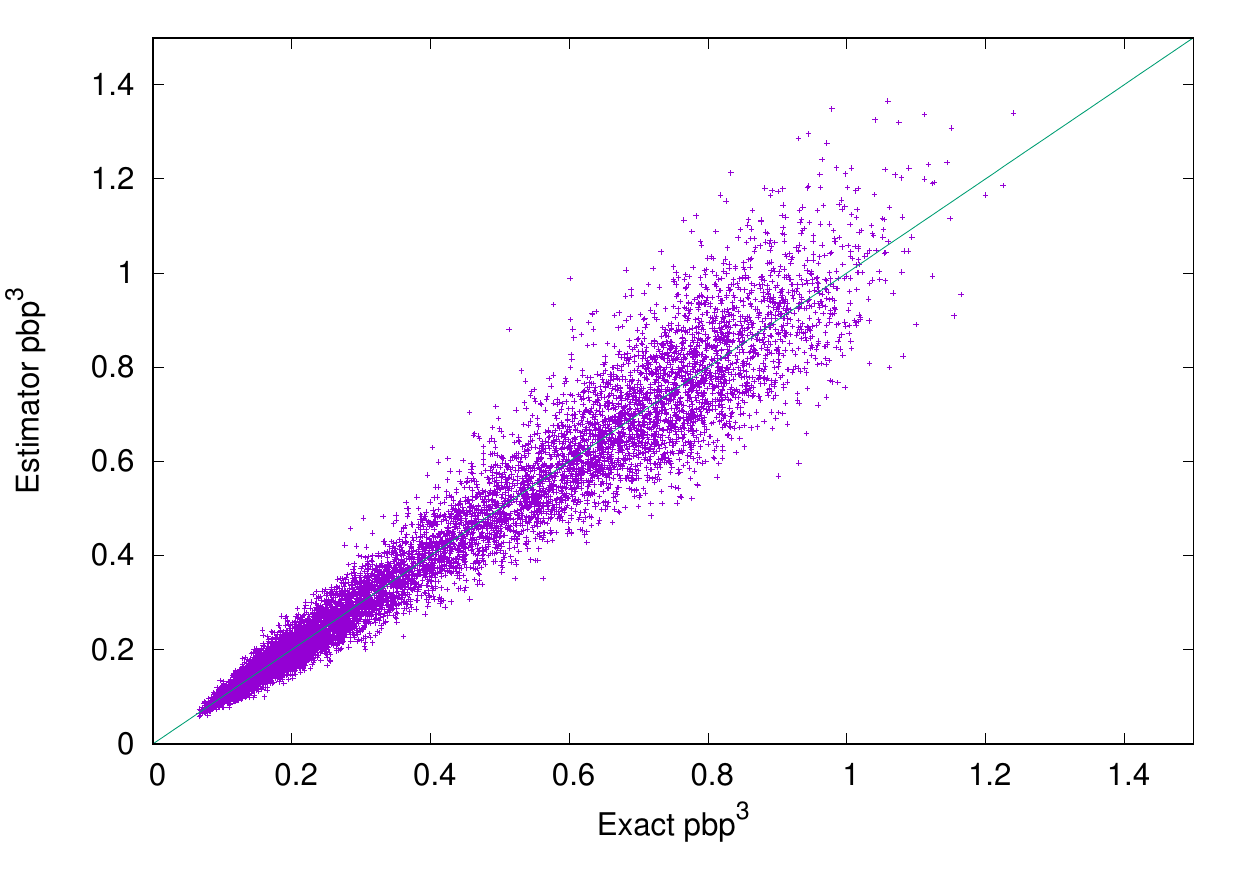}
\includegraphics[width=0.50\linewidth]{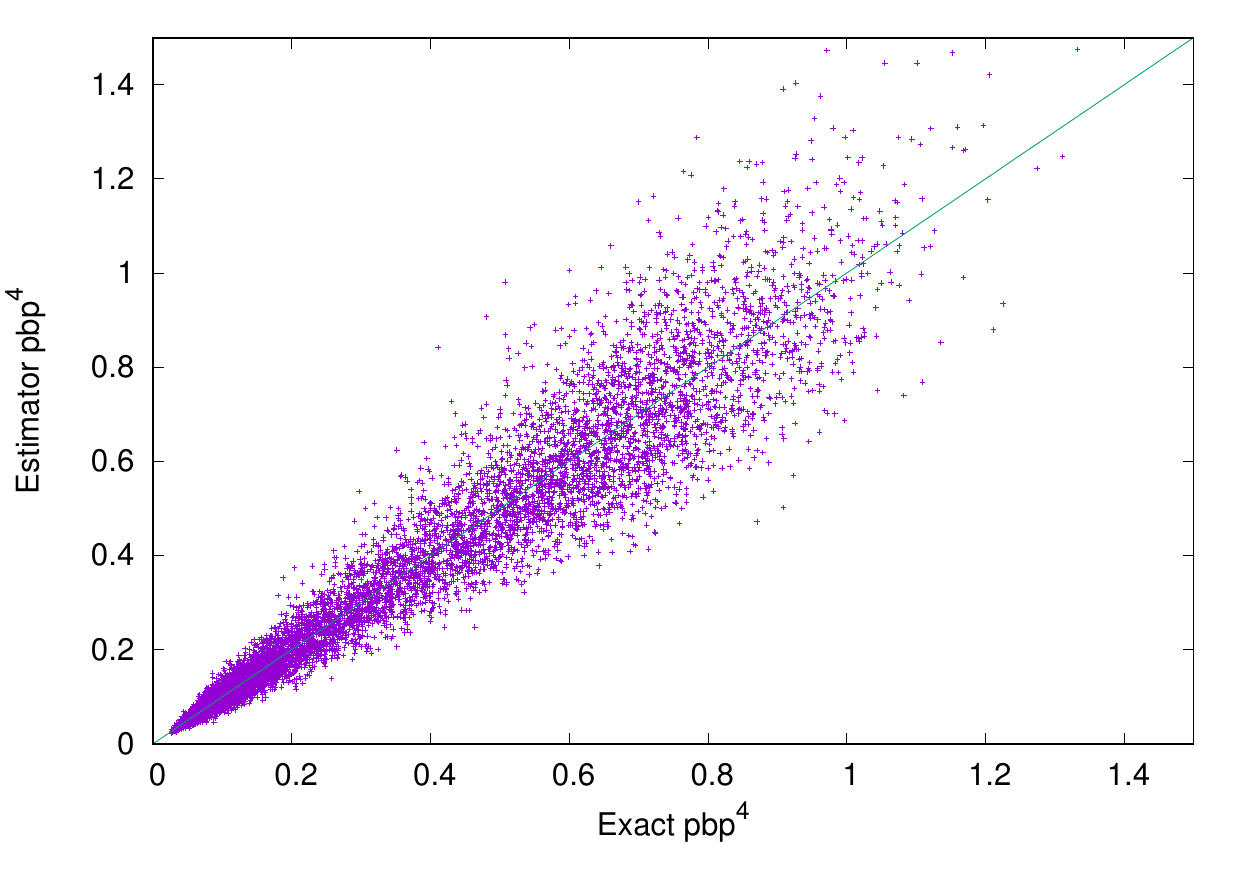} \\
\hspace*{3.5cm}
\includegraphics[width=0.50\linewidth]{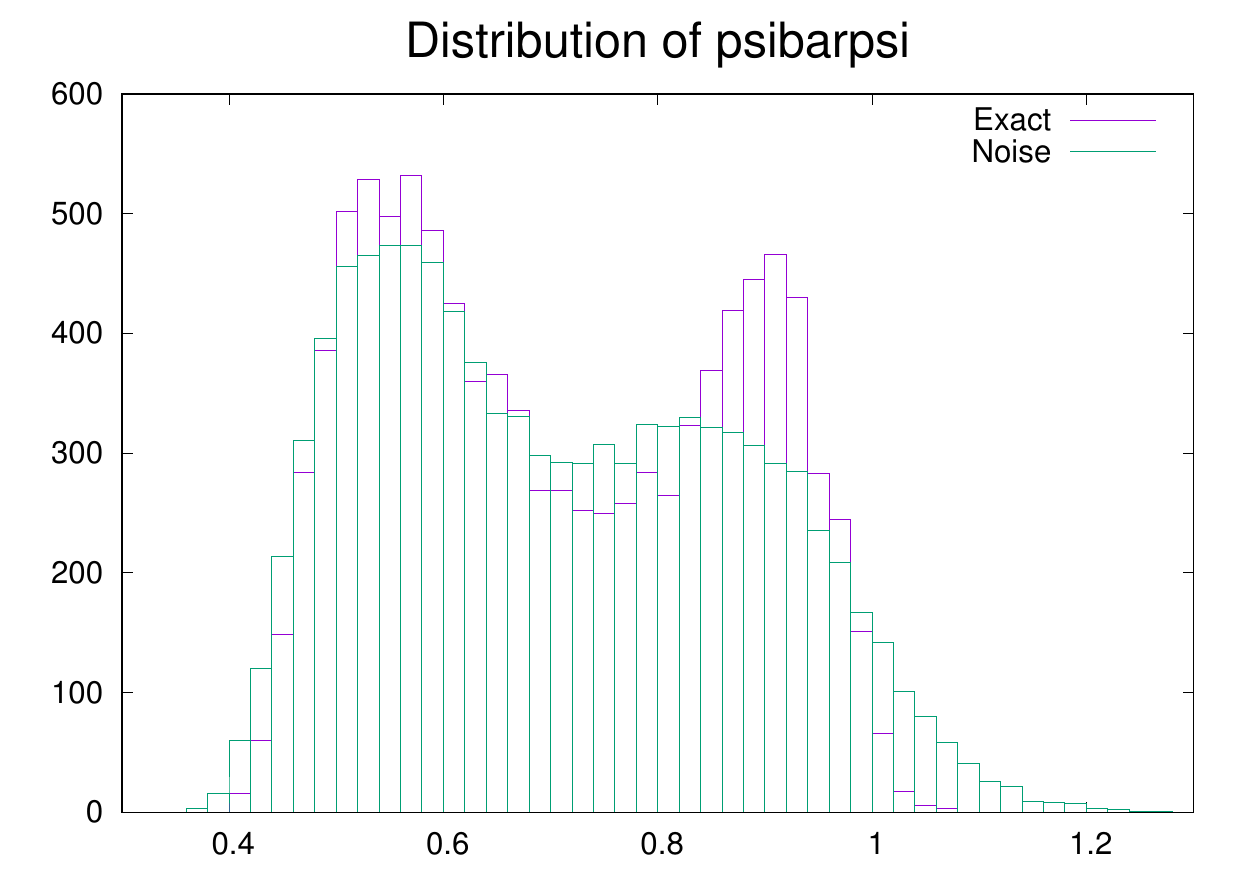}
\caption{Comparison of the first 4 powers of $\bar\psi \psi$ with their 
stochastic estimators eq.(\ref{eq_3}), on an ensemble of $6^3\times 4$
SU$(3), N_f=4$ configurations with staggered fermions, near the
finite-temperature deconfinement transition. 
The bottom figure compares the resulting distribution of the exactly
measured $\bar\psi \psi$ with that of its estimator: the two-bump
distribution signals the vicinity of the finite-$T$ transition; the "stochastic
noise" barely increases the variance of the distribution. }
\end{figure}

%---------------------------------------------------------------------------

\section{Application I: Binder cumulant for the finite-temperature transition}\label{sec-3}

A suitable order parameter for the finite-$T$ transition is the quark condensate $\bar\psi \psi$.
To probe the order of the phase transition, it is customary to measure its Binder cumulant
(or equivalently its kurtosis) 

\be
B_4(\bar\psi \psi) \equiv \frac{\langle (\delta\bar\psi \psi)^4 \rangle}{\langle (\delta\bar\psi \psi)^2 \rangle^2}, ~~~~ \delta\bar\psi \psi = \bar\psi \psi - \langle \bar\psi \psi \rangle 
\ee

\noindent
Thus, the first 4 moments $(\bar\psi \psi)^k, k=1,..,4$ must be measured, or estimated in an unbiased way.
The latter approach is normally chosen, using noise vectors. One must then decide how many noise vectors
per configuration to use: using too few will increase the ``stochastic noise'' on $B_4$ and the associated
statistical error; using too many is wasteful of computing resources. \\
Here, we chose the most economical approach: 4 noise vectors $\eta_1,..,\eta_4$, forming unbiased 
estimators as
\bea
\bar\psi \psi \phantom{)^2} & \longleftarrow & \frac{1}{4} \left( \eta_1^\dagger \Dslash^{-1} \eta_1 + \eta_2^\dagger \Dslash^{-1} \eta_2 + \ldots \right) \nn \\ 
(\bar\psi \psi)^2 & \longleftarrow & \frac{1}{6} \left( \eta_1^\dagger \Dslash^{-1} \eta_1 \cdot \eta_2^\dagger \Dslash^{-1} \eta_2 + \ldots \right) \nn \\
(\bar\psi \psi)^3 & \longleftarrow & \frac{1}{4} \left( \eta_1^\dagger \Dslash^{-1} \eta_1 \cdot \eta_2^\dagger \Dslash^{-1} \eta_2  \cdot \eta_3^\dagger \Dslash^{-1} \eta_3 + \ldots \right) \nn \\
(\bar\psi \psi)^4 & \longleftarrow & \left( \eta_1^\dagger \Dslash^{-1} \eta_1 \cdot \eta_2^\dagger \Dslash^{-1} \eta_2 + \eta_3^\dagger \Dslash^{-1} \eta_3 \cdot \eta_4^\dagger \Dslash^{-1} \eta_4 \right) 
\label{eq_3}
\eea

In Fig.~1, we compare the estimators above with the exact values obtained by diagonalization of $\Dslash$, for an ensemble of $6^3\times 4$ SU$(3)$ configurations, where
the parameters ($N_f=4, am=0.07$) have been chosen to be in the vicinity of a sharp temperature-driven crossover. Indeed, the exact distribution of $\bar\psi \psi$ in the
last panel shows a clear two-peak distribution. What is striking is that the distribution of the stochastic estimator of $\bar\psi \psi$, in the same figure, is hardly broader,
even though the number of noise vectors has been kept to its minimum value 4. Thus, in this example, the ``stochastic noise'' is negligible compared to the
``gauge noise''. Whether or not this persists on larger lattices needs to be checked.

\begin{figure}[thb] 
  \centering
  \includegraphics[width=0.45\linewidth]{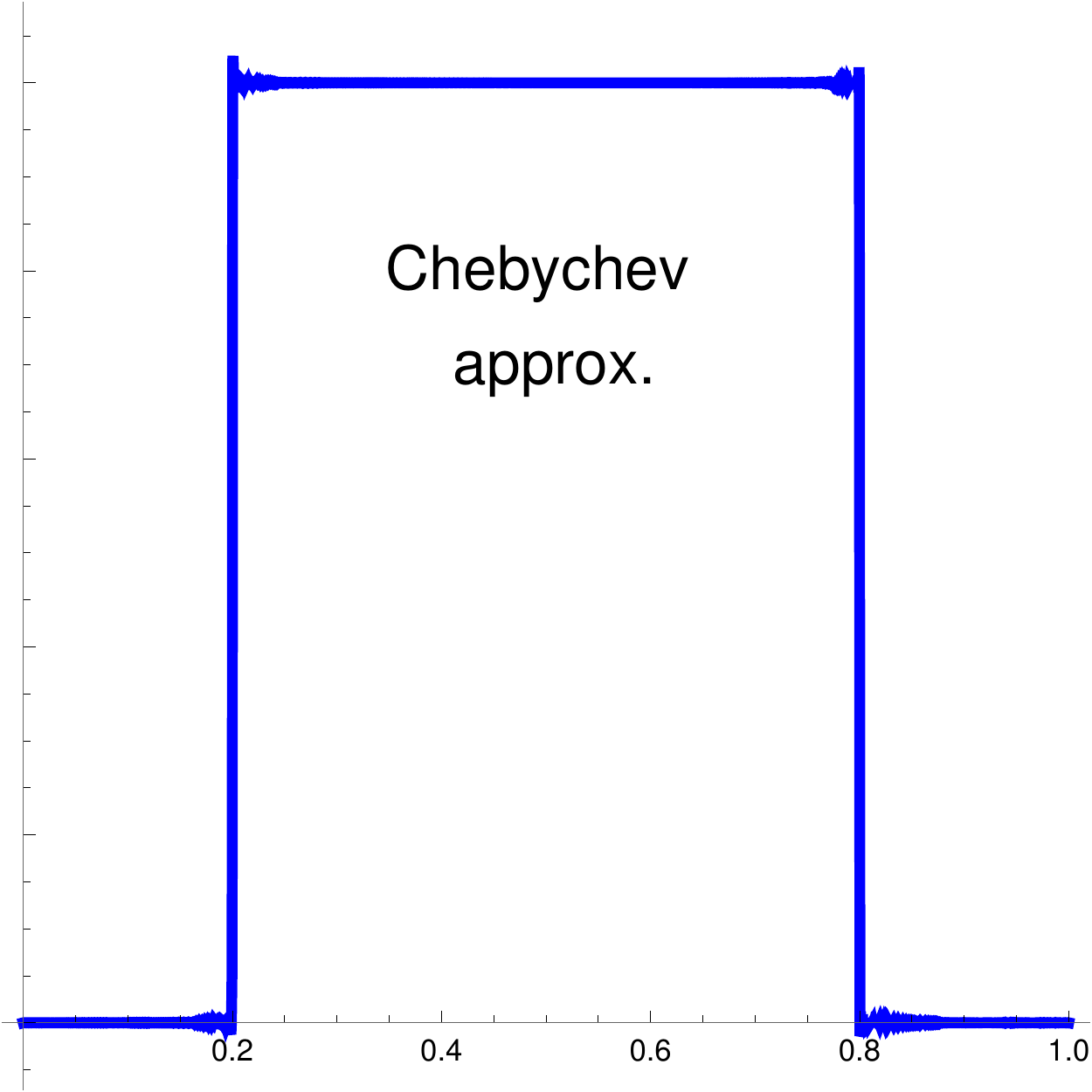}
  \hspace*{1.0cm}
  \includegraphics[width=0.45\linewidth]{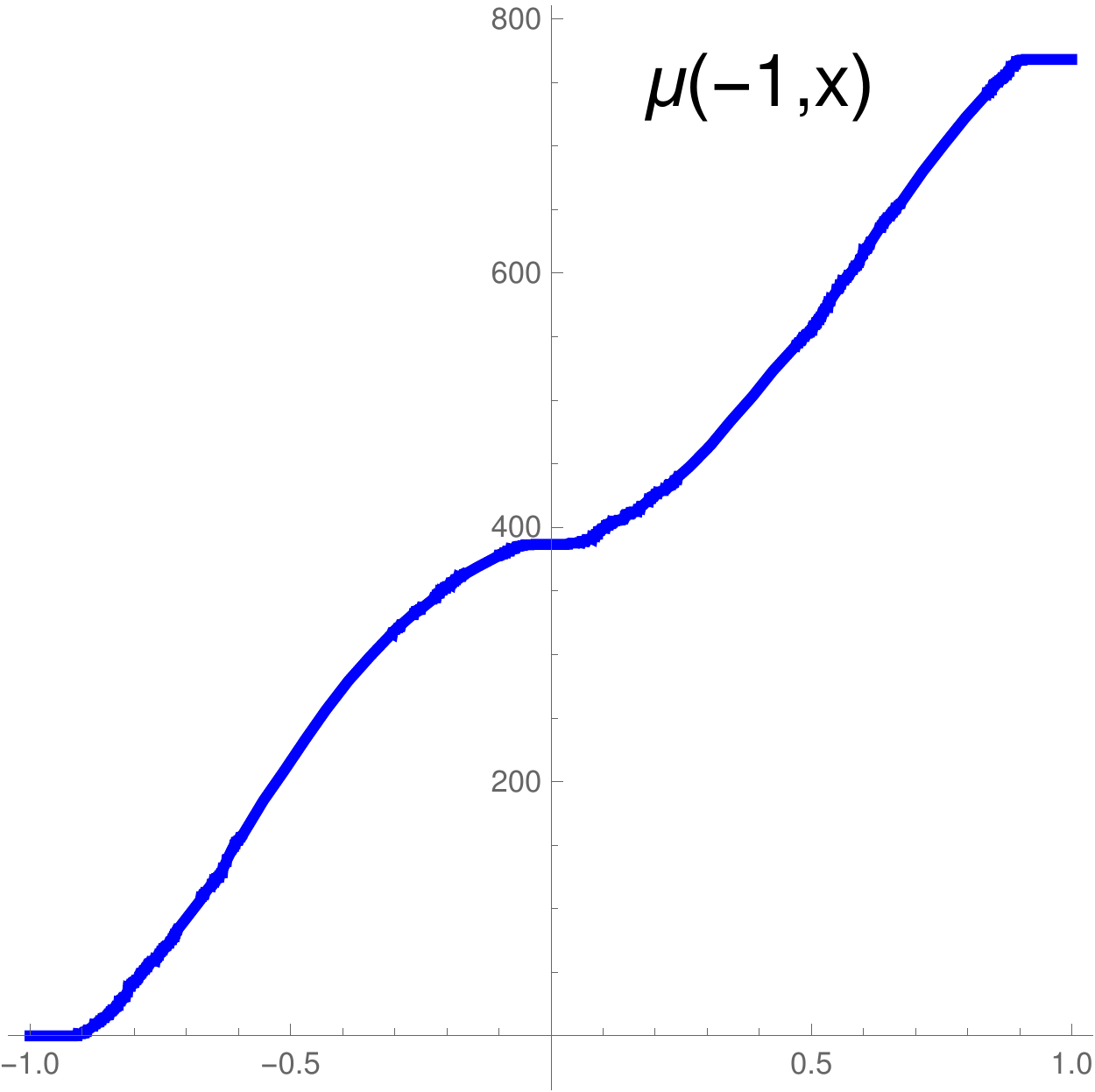}
  \caption{Chebyshev bandpass filter ({\em left}) and eigenvalue count $\mu(-1,x)$ as a function of $x$, for the rescaled, staggered
operator $(i \Dslash)$ on a $4^4$ lattice. }
  \label{fig-2}
\end{figure}

The novel method outlined in Sec.~2 can give us nearly exact results at a potentially reduced cost, compared with exact diagonalization.
A Chebyshev bandpass filter is applied to $(i \Dslash)$, so that the number $\mu$ of eigenvalues in $[a,b]$ can be approximated as

\be
 \mu (a,b) = \frac{1}{n} \sum_{j=1}^{n} \sum_{p=0}^{p_\text{max}} g_p(a,b) \,
\langle \langle \eta_j^\dagger T_p(i  \Dslash) \, \eta_j
\rangle\rangle
\ee

\noindent
where $T_p$ is the Chebyshev polynomial of degree $p$, $g_p$ are the coefficients of the filter approximation, and $\langle\langle \cdots \rangle\rangle$ 
represents the averaging over the gauge field. Fig.~2 illustrates the bandpass filter ({\em left}) and the eigenvalue count $\mu(-1,x)$ versus $x$ ({\em right})
\footnote{Note the similarity with L\"uscher's ``mode
number''~\cite{Giusti:2008vb}.}, for one $4^4$ configuration ($\Dslash$
has been rescaled so that its spectrum is contained in $[-1,+1]$). $\mu(-1,x)$ is a piecewise constant function, increasing by one as $x$ passes each eigenvalue. We approximate the eigenvalues as solutions of  $\mu(-1,x)$ being half-integer.
Since all eigenvalues are obtained in this way, we can form arbitrary moments $\Tr\left( \Dslash^{-k} \right), k=1, 2, \ldots$. 
This is illustrated in Table~1 below. Note the remarkable accuracy, even for high moments.

\begin{table}[!htb] 
\begin{center}
\begin{tabular}{|c|c|c|c|}
        \hline
        $k$ & exact  & approx. & approx./exact \\
        \hline
        $2$ & $-453.767$ & $-451.148$ & $0.994229$ \\
        $4$ & $3871.66$ & $3844.25$ & $0.99292$ \\
        $10$ & $-8.136\cdot10^7$ & $-8.091\cdot10^7$ & $0.994547$ \\
%        $\frac{\Tr(\Dslash^{-4})}{\Tr(\Dslash^{-2})^2}$ & $0.0188032$ & $0.0186701$ & $1.00713$  \\
        \hline
\end{tabular}
\end{center}
\caption{$\Tr \left(\Dslash^{-k} \right)$ using 4 noise vectors
and a degree-16k polynomial on a $4^4$ lattice.}
\end{table}

%---------------------------------------------------------------------------

\section{Application II: Taylor expansion of free energy}\label{sec-4}

QCD at non-zero quark chemical potential $\mu$ represents a long-standing challenge for lattice simulations:
the Dirac determinant becomes complex when $\mu\neq 0$, which prevents its interpretation as a sampling
probability. Several strategies have been proposed to circumvent this ``sign'' (or phase) problem.
One of the first is that of Taylor-expanding the QCD free energy, or the
pressure $P$, in powers of $(\mu/T)$ around $\mu=0$:
\be
P(\mu,T) = P(\mu=0,T) + \sum_{k=1} c_k(T)  \left( \frac{\mu}{T} \right)^k.
\ee
Charge conjugation symmetry implies that odd-order coefficients 
$c_k$ vanish. However, products of odd derivatives do contribute in the presence of multiple quark flavors, each with its chemical potential.
Now, $c_k$ is simply the $k^{th}$ derivative of the free energy evaluated at $\mu=0$, so it can
be measured in conventional, sign-problem free simulations. 
The $\mu$-dependence appears only in the Dirac determinant, so that the generic expression for
$c_k$ is the trace of a polynomial expression in $\Dslash^{-1}$ and in derivatives $\partial^m \Dslash/\partial \mu^m$.
The number of terms grows exponentially with the order $k$, with positive or negative signs.
This makes the calculation of high-order coefficients $c_k$ extremely challenging: no calculation
has ever yielded information at $10^{th}$ order. Improvements in efficiency are needed.

One such improvement has been proposed
recently by Gavai and Sharma~\cite{Gavai:2014lia,Gavai:2015ywa}.
The suggestion is to change the way the chemical potential is introduced on the lattice, and adopt a naive, linear
prescription rather than the exponential prescription of Hasenfratz and Karsch~\cite{hasenfratz1983chemical}.
This is rather counter-intuitive, since it re-introduces a UV divergence in the pressure.
Gavai and Sharma argue that this divergence can be removed by a numerical subtraction.
The gain is that derivatives $\partial^m \Dslash/\partial \mu^m$ now vanish for $m > 1$.
This greatly simplifies the structure of the $c_k$'s, which become the traces of homogeneous polynomials of
degree $k$ in $A^{-1}$, with $A = \Dslash \left( \frac{\partial \Dslash}{\partial \mu} \right)^{-1}$
(note that $A$ is neither hermitian, nor anti-hermitian -- see Fig.~3).

At this stage, our approach may become advantageous. Suppose we can determine the [complex] spectrum
of $A$. Then, $c_k$ will be obtained as a sum of products of factors $\Tr \left( A^{-m} \right)$, which can
{\em all} be determined in one go!

Here, we take first steps in that promising direction. 
First, we generalize the notion of eigenvalue count to the complex plane.
The number of eigenvalues $\mu(\Gamma)$ inside a complex contour $\Gamma$ can be obtained
by a Cauchy integral~\cite{futamura2010parallel}:
\be
\mu(\Gamma) = \frac{1}{2\pi \ii} \oint_\Gamma  dz ~  \Tr  \left[ \left( z \id
- A \right)^{-1} \right]  = \frac{1}{2\pi \ii} \oint_\Gamma dz ~ \sum_j
\frac{1}{z -\lambda_j}
\ee
For $\Gamma$, we choose a circle of center $c$ and radius $r$, and discretize the integral using the
trapezoidal rule:
\be
\mu(r,c) \sim \frac{1}{N} \sum_{k=0}^{N-1} r \, \text{e}^{\frac{2 \pi \ii}{N}
k }\,\, \Tr  \left[ \left( z_k \id - A
\right)^{-1} \right] \text{ with } z_k= c + r \, \text{e}^{\frac{2 \pi \ii}{N}  k }
\label{eq_7}
\ee
where the trace is estimated using noise vectors.
The so-obtained value of $\mu(r,c)$ will be nearly an integer equal to the number of eigenvalues
inside the circle. One can now discard the circles which contain zero eigenvalue, and divide by 2 the
radius of non-empty circles. This procedure is repeated recursively, until each circle contains
one and only one eigenvalue, and has a radius smaller than some prescribed tolerance $\epsilon$.
In this fashion, the complete spectrum of $A$ is determined. Arbitrary moments $\Tr A^{-m}$ can
be evaluated.

The crucial point is that the estimation of the trace in eq.(\ref{eq_7}) proceeds by solving
linear systems $\left( z_k \id - A \right) x = \eta$, using a multi-shift linear solver. Then the cost of solving many such systems is not
much greater than that of solving just one, because these systems differ only by the shift $z_k$,
and the Krylov space built by the linear solver is shift-invariant. 
In addition, each iteration of the solver involves matrix-vector multiplication by $A =  \Dslash \left( \frac{\partial \Dslash}{\partial \mu} \right)^{-1}$.
At first sight, the inversion of $\frac{\partial \Dslash}{\partial \mu}$ looks daunting. But this operator is completely local in space, so that
the corresponding matrix is made of small diagonal blocks, each of which is easy to invert.

We illustrate our approach in Figs.~3 and 4. 
Fig.~3 shows the complex spectrum of $A$, for 3 configurations obtained with the same action, near the
finite-$T$ transition, on lattices of increasing spatial sizes.

\begin{figure}[t] 
  \centering
\includegraphics[width=0.32\linewidth]{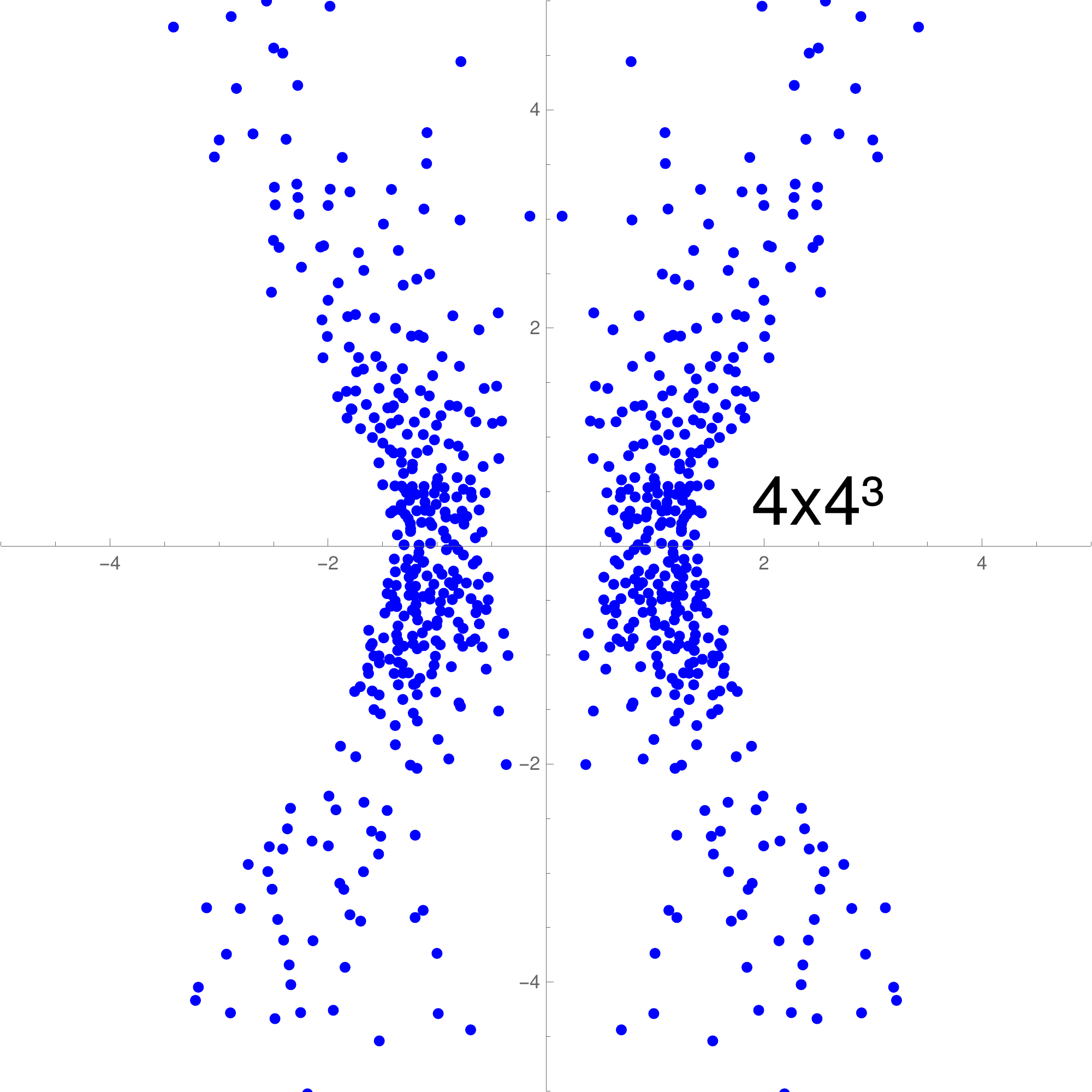}
\includegraphics[width=0.32\linewidth]{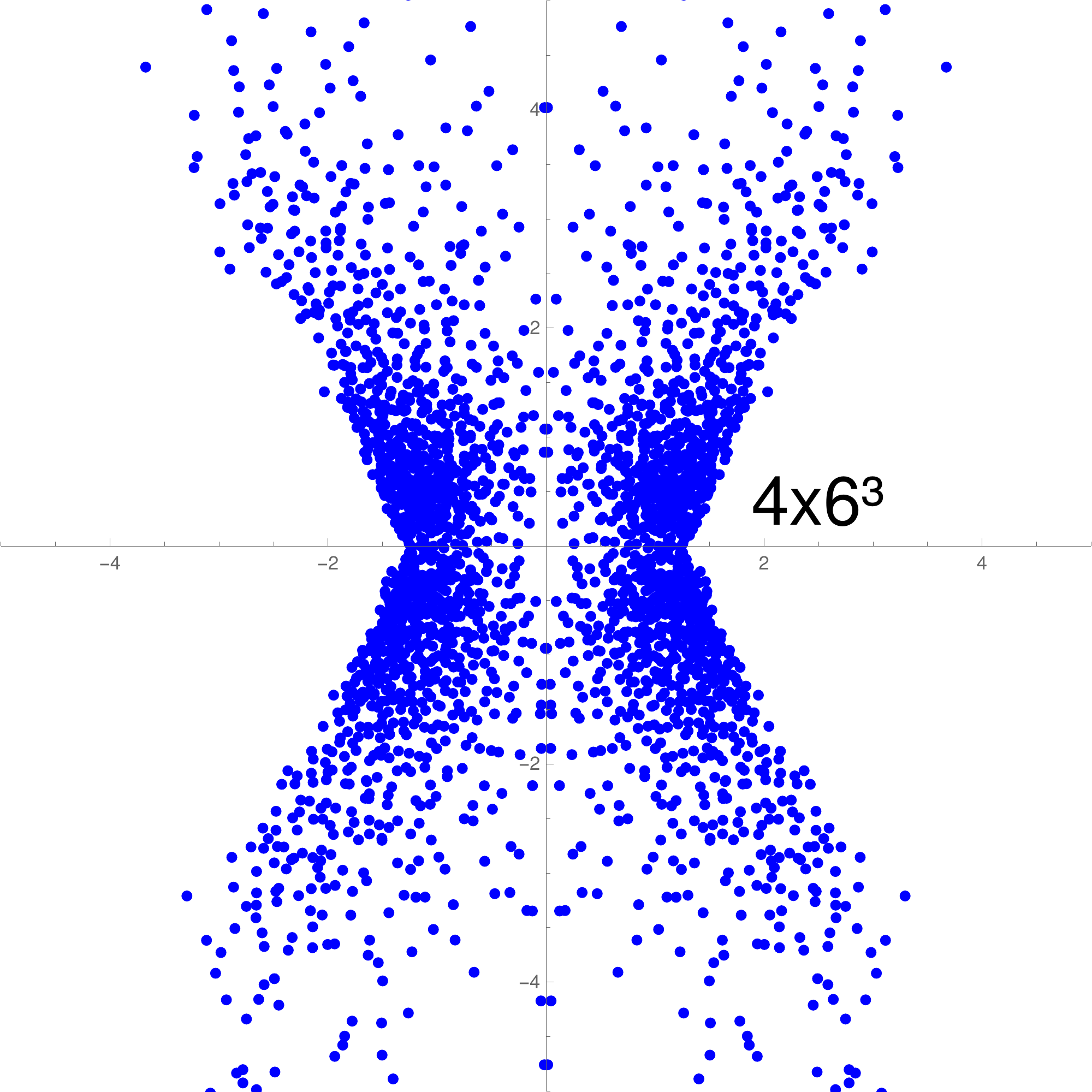}
\includegraphics[width=0.32\linewidth]{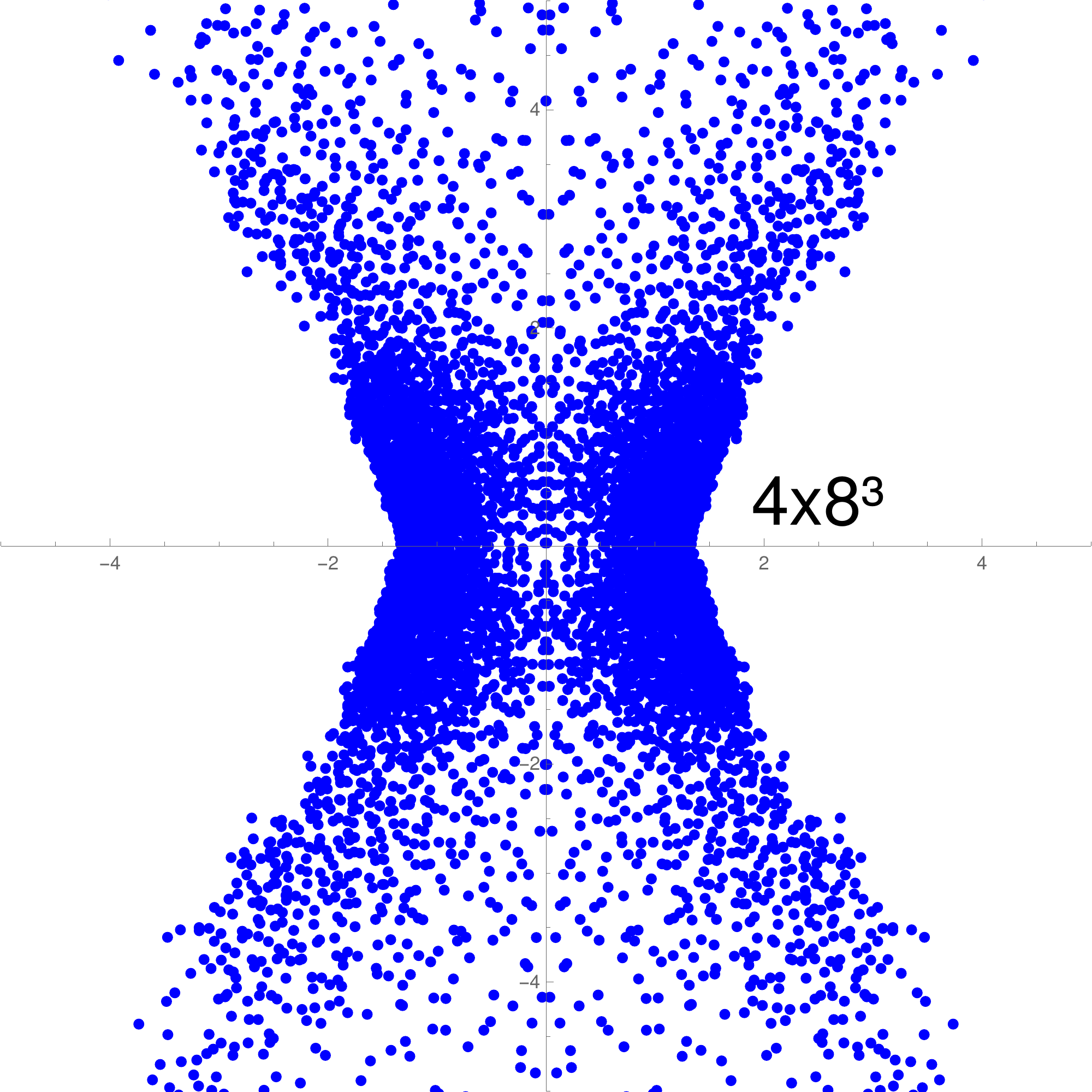}
 \caption{Complex spectrum of $\left( \Dslash  \left( \frac{\partial \Dslash}{\partial \mu}\right)^{-1}  \right)$, for lattices of increasing spatial size near the finite-$T$ transition.}
  \label{fig_3}
\end{figure}

\begin{figure}[t] 
  \centering
\includegraphics[width=0.32\linewidth]{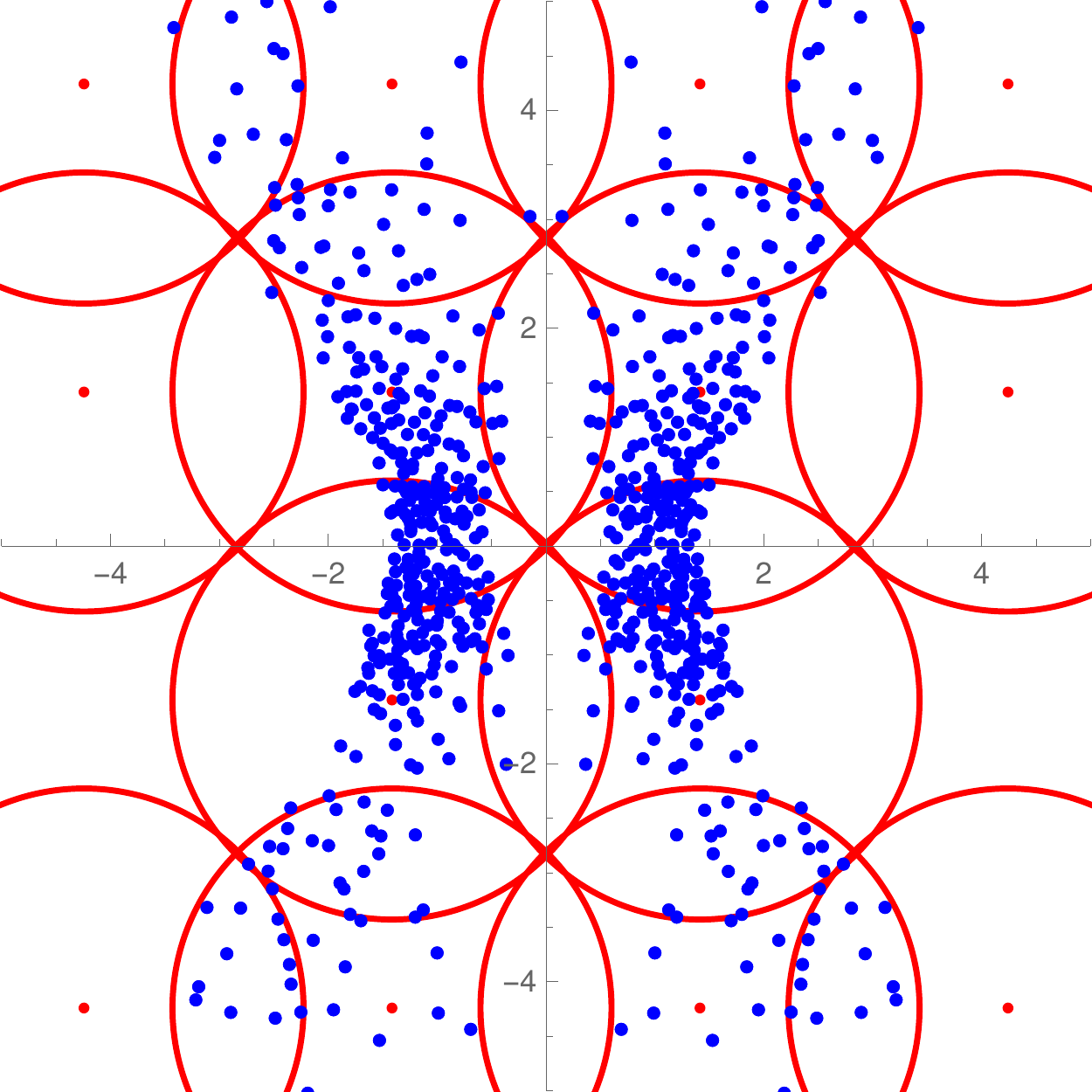}
\includegraphics[width=0.32\linewidth]{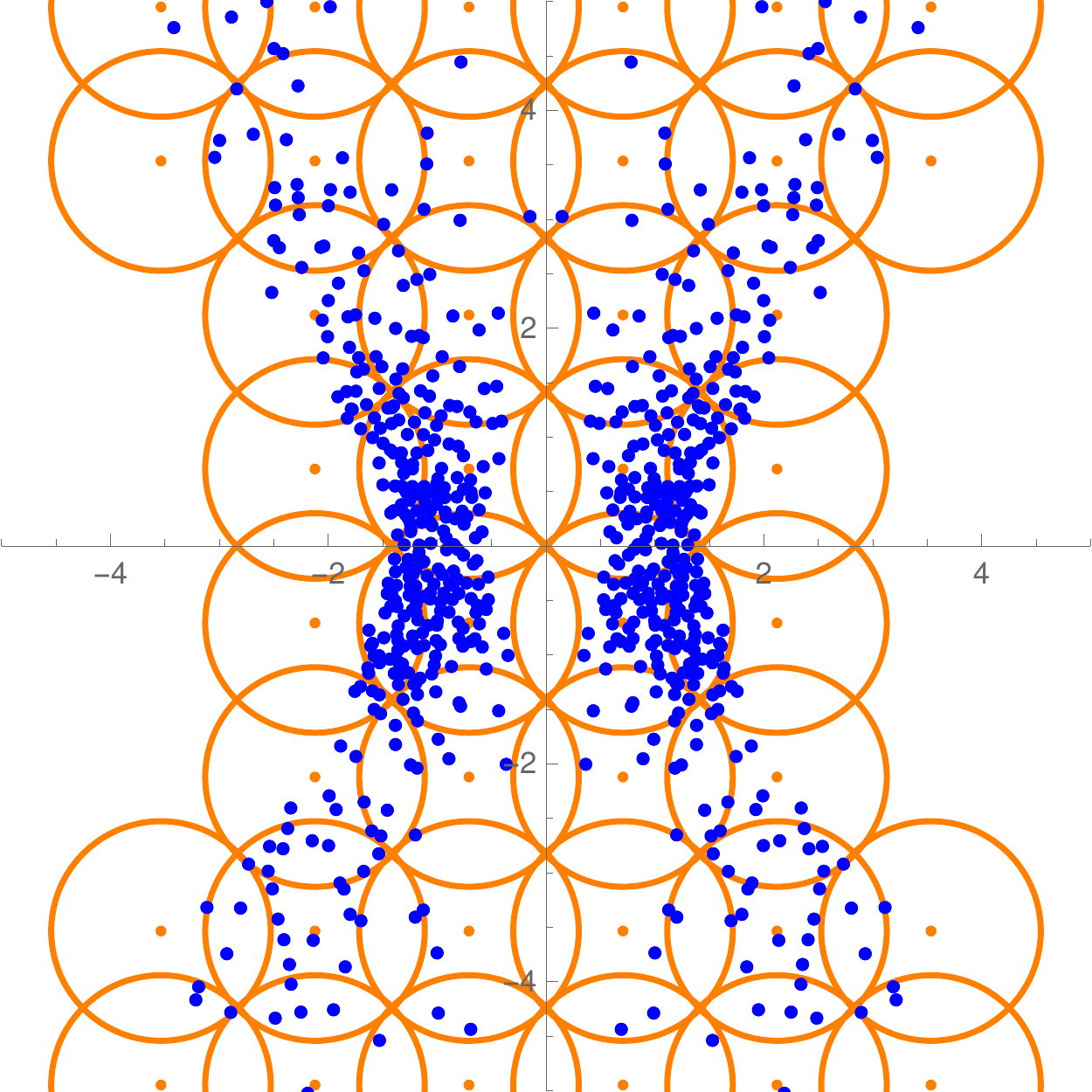}
\includegraphics[width=0.32\linewidth]{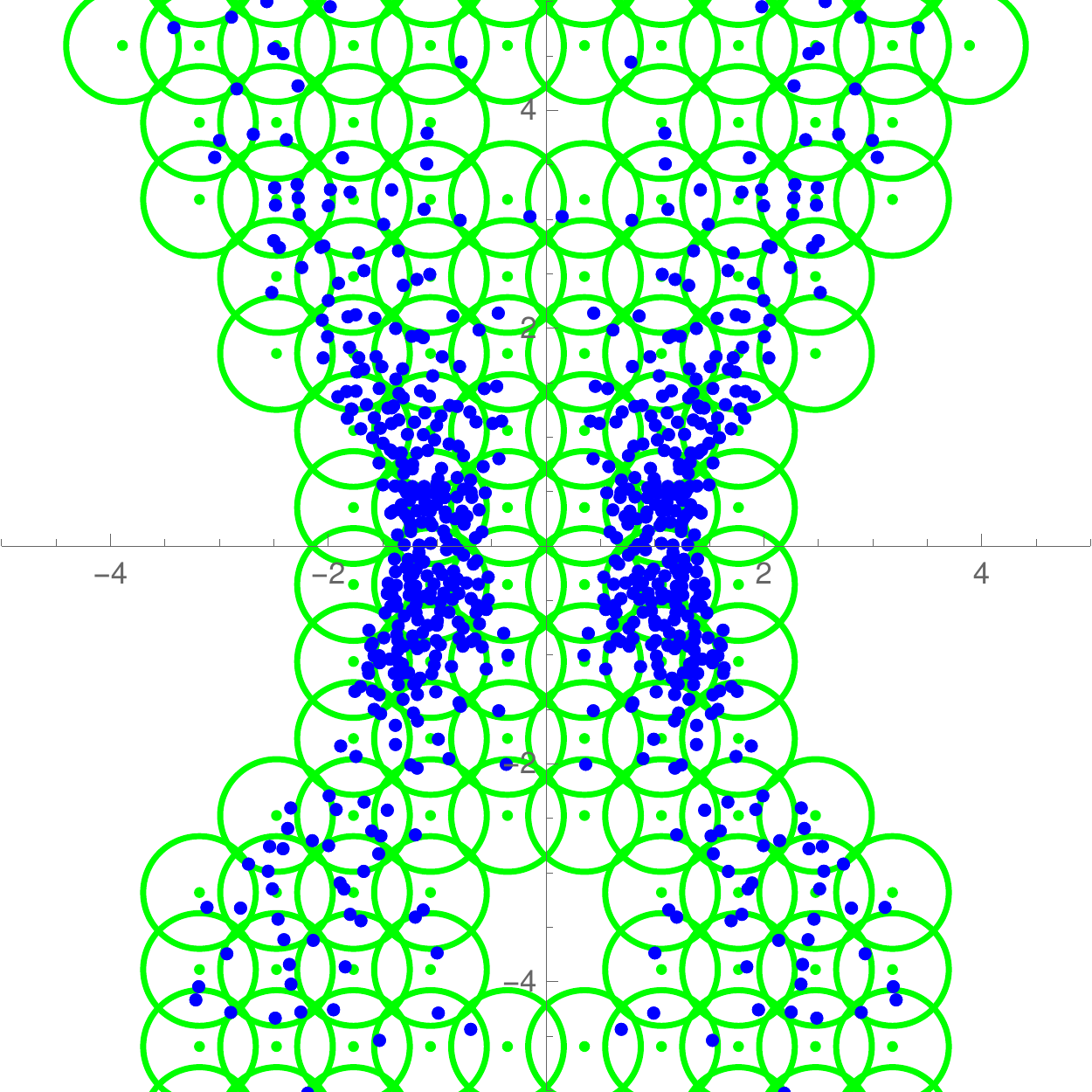}
 \caption{Circular contours are recursively refined until each contour encloses one eigenvalue (or a few).}
  \label{fig_4}
\end{figure}

Fig.~4 shows 3 successive stages of our recursive procedure: at each stage, circles empty of eigenvalues are discarded, and 
non-empty circles are replaced by 4 circles with half the radius. The
final output is an estimate of all eigenvalues of $A$, from which successive moments $\Tr \left( \Dslash \, \frac{\partial \Dslash}{\partial \mu}^{-1} \right)^{-m}$ can be estimated as well.
Illustrative results are provided in Table~2 below. 

\begin{table}[!htb]
\begin{center}
\begin{tabular}{|c|c|c|c|}
        \hline
        m & exact & approx. & ratio \\
        \hline
        $1$ & $4.10306\ii$ & $0.0352058 + 4.10474\ii$ & $1.00041$ \\
        $2$ & $142.497$ & $142.607 - 0.00748385\ii$ & $1.00077$ \\
        $3$ & $10.8641\ii$ & $0.04117+10.8765\ii$ & $1.00114$ \\
        $4$ & $-14.4443$ & $-14.3171 - 0.0751784\ii$ & $0.991195$ \\
        $10$ & $-133.716$ & $-135.113 - 3.93177\ii$ & $1.01045$ \\
        $20$ & $-79320.5$ & $-81511.3 + 979.464\ii$ & $1.02762$ \\
        \hline
\end{tabular}
\end{center}
\caption{$\Tr \left(A^{-m} \right)$ for 100 quadrature points and
100 noise vectors for $\epsilon=10^{-3}$, on a $4^4$ gauge configuration.}
\end{table}

\section{Outlook}

We have described a novel approach to obtain estimates of moments $\Tr \left(A^{-m} \right)$
of a sparse matrix $A$, in the Hermitian and in the non-Hermitian case.
Two applications were tested: measuring the Binder cumulant of the quark condensate ($A = i \Dslash$),
and the coefficients of the Taylor expansion of the pressure in $\mu/T$ ($A = \Dslash \left( \frac{\partial \Dslash}{\partial \mu} \right)^{-1}$).
In both cases, the regime of interest is near the finite-temperature deconfinement transition.
In this regime, fluctuations of the gauge field are large, so that the ``gauge noise'' may well exceed the ``stochastic noise'' caused by the
limited number of noise vectors used to estimate $\Tr \left(A^{-m} \right)$. 
Indeed, this is what happens in the first application: it is then sufficient to use 4 noise vectors per configuration, at least for
the small lattice sizes considered in this test.

In the second application, we have generalized the eigenvalue count to the complex plane.
The viability of our approach rests on a massive usage of non-Hermitian multi-shift linear solvers.
Great simplifications in the contour integrals are possible and are under study.
However, while we obtain nearly exact results for the first 20 moments $\Tr \left(A^{-m} \right)$, we have not 
measured the ``gauge noise'', which presumably increases quickly with $m$. This may represent the
ultimate, irreducible origin of the computing cost of the Taylor expansion of the pressure, as foreseen in~\cite{deForcrand:2010ys}.

\begin{textblock}{0}(5.15,3.0)
{\Large $\Rightarrow$ }
\end{textblock}
\begin{textblock}{0}(9.5,3.0)
{\Large $\Rightarrow$ }
\end{textblock}

\section{Acknowledgements}\label{Acknowledgements}
BJ was supported by the Schweizerischer Nationalfonds
(SNF) under grant 200020-162515.
PdF thanks the Kavli Institute for Theoretical Physics for hospitality.
This research was supported in part by the National Science Foundation 
under Grant No. NSF PHY-1125915.

%\clearpage
\bibliography{lattice2017}

%%%%%%%%%%%%%%%%%%%%%%%%%%%%%%%%%%%%%%%%%%%%%%%%%%%%%%%%%%%%%%%%%%%%%%%%%%%%%
\end{document}